\documentclass[conference,a4paper]{IEEEtran}
\IEEEoverridecommandlockouts
\usepackage{cite}
\usepackage{tikz}
\usetikzlibrary{arrows.meta, positioning, calc}
\usepackage{amsmath, amssymb}
\usepackage[T1]{fontenc}
\usepackage{helvet}
\usepackage{amsmath,amssymb,amsfonts}
\usepackage{algorithmic}
\usepackage{graphicx}
\usepackage{textcomp}
\usepackage{xcolor}
\usepackage[utf8]{inputenc}
\usepackage[T1]{fontenc}
\usepackage{cite}
\usepackage{amsmath,amssymb,amsfonts}
\usepackage{algorithmic}
\usepackage{graphicx}
\usepackage{textcomp}
\usepackage{xcolor}
\usepackage{booktabs} 
\usepackage{multirow} 
\usepackage{url}      
\usepackage{bm}       
\usepackage{algorithm}      
\usepackage{algorithmic}    
\def\BibTeX{{\rm B\kern-.05em{\sc i\kern-.025em b}\kern-.08em
    T\kern-.1667em\lower.7ex\hbox{E}\kern-.125emX}}
\begin{document}

\title{Cluster-Aware Dual-Level Test Specification Generation for Large-Scale Automotive Software Requirements}

\author{\IEEEauthorblockN{Hazem Ayman, Menna Sedik, Kareem Mostafa, Mahmoud Soliman*, Samer Saber, Ibrahim Habib}
\IEEEauthorblockA{\textit{CairoMotive} \\
Cairo, Egypt \\
\textsuperscript{*}mahmoud.soliman@cairomotive.com}
}


\maketitle

\begin{abstract}
Generating test specifications that satisfy Automotive SPICE SWE.6 requirements becomes increasingly challenging and time-consuming as projects scale to thousands of requirements. Because this manual process often consumes weeks of engineering effort, automation becomes a critical necessity. However, standard Large Language Model (LLM) approaches struggle at scale: processing requirements individually discards vital inter-requirement dependencies, while feeding entire corpora at once exceeds context-window limits, leading to incomplete integration coverage and redundant test cases. This paper presents a novel "Cluster-then-Summarize" pipeline that addresses these limitations through three-stages. Requirements are embedded using sentence transformers and grouped using UMAP dimensionality reduction followed by HDBSCAN density-based clustering. This grouping utilizes an automatic minimum cluster size selection driven by a quality criterion combining normalized Silhouette and Calinski–Harabasz scores. A multi-level map-reduce summarization algorithm then distills each cluster into concise, domain-conformant descriptions while preserving quantitative thresholds and safety integrity levels. The pipeline exploits the derived cluster topology to generate test specifications at two levels: individual requirement verification and cluster-level integration tests that verify cross-requirement feature behavior. A nearby-cluster context mechanism provides bounded cross-feature awareness during each LLM call, and Retrieval-Augmented Generation grounds all outputs in ISO~26262 and ASPICE standards. Evaluation on automotive requirement datasets of varying scale demonstrates that the cluster-aware approach improves integration test coverage and maintains summarization fidelity compared to baseline methods while scaling efficiently to thousands of requirements.
\end{abstract}


\section{Introduction}
Software Requirements Engineering (SRE) is widely recognized as the most critical phase of the Software Development Life Cycle, where errors and ambiguities introduced early propagate exponentially into design, implementation, and testing \cite{necula2024slr, rosadadacruz2025ml}. In safety-critical domains such as automotive software development, this challenge is amplified by strict regulatory frameworks. Automotive SPICE (ASPICE) defines process areas governing the entire development lifecycle, with SWE.6 specifically mandating that every software requirement be verified through traceable test specifications \cite{aspice2017}. Compliance with functional safety standards such as ISO~26262 further requires that test coverage account for Automotive Safety Integrity Levels (ASILs) and inter-requirement dependencies \cite{iso26262}.
\begin{figure}[t]
    \centering
    \includegraphics[width=\linewidth]{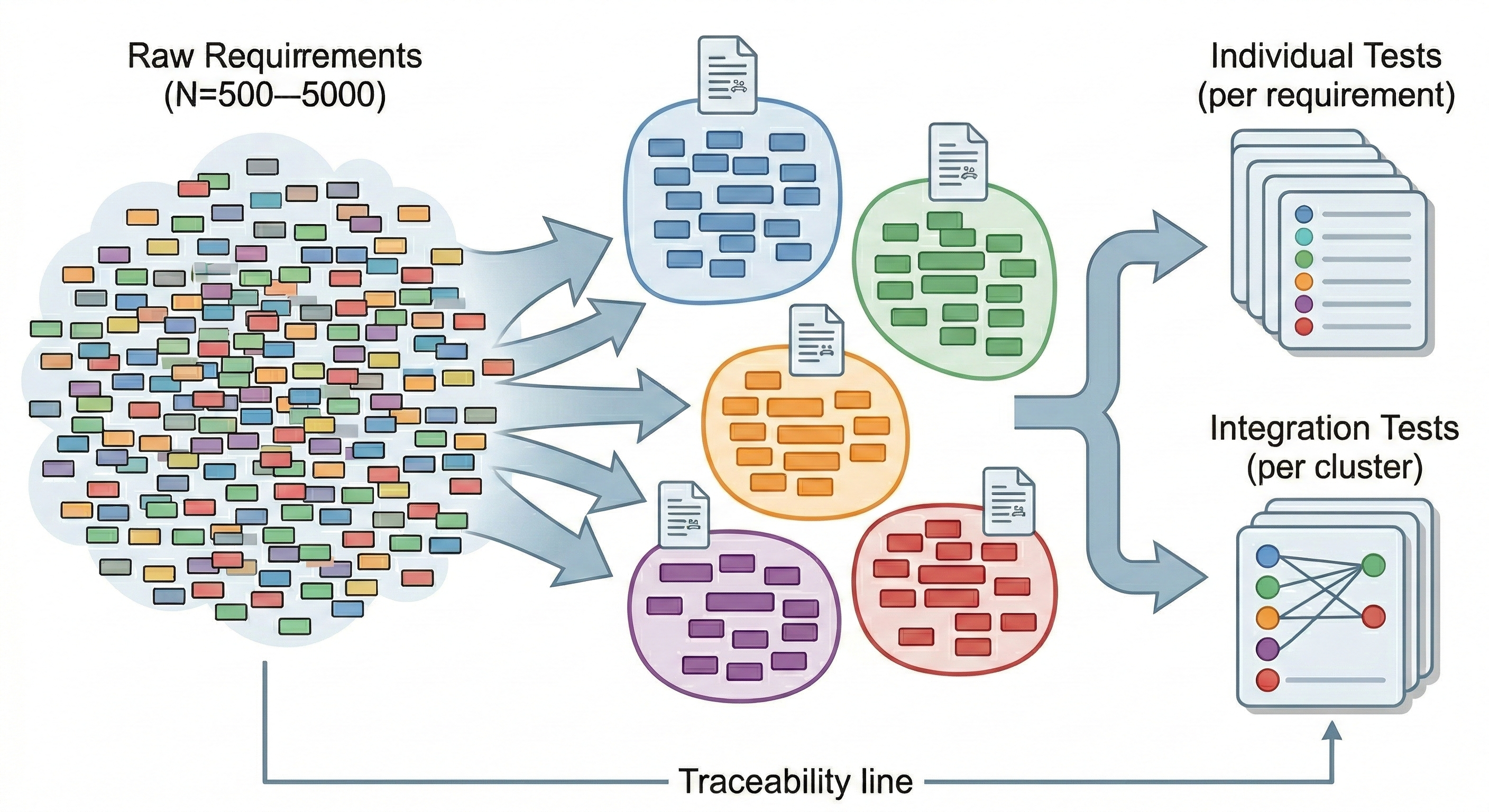} 
    \caption{Requirement engineering workflow for converting raw data into clustered, traceable test cases.}
    \label{fig:erm_failure}
\end{figure}
A typical automotive project includes 500 to 5{,}000+ requirements across domains such as powertrain, ADAS, and V2X. Producing complete test specifications is still largely manual and requires both requirement-level understanding and cross-requirement reasoning over large unstructured documents \cite{spijkman2023}, as illustrated in Fig.~\ref{fig:erm_failure}.

NLP4RE reduces this manual burden \cite{nlp4re2020mapping}. With modern LLMs, the field has moved from rule-based mining to semantic analysis \cite{llm4re2025}. Two core techniques are summarization, which condenses large requirement sets \cite{abstractive2024survey}, and clustering, which groups semantically related requirements \cite{rosadadacruz2025ml}. Prior work shows that combining clustering and summarization scales well for multi-document requirements corpora \cite{chatdialogue2021, hierarchicalsumm2014}.

In SRE, clustering supports redundancy detection, document structuring, and conflict detection \cite{sharma2025kmeans, clusterstructure2014, conflictdetection2022}. SBERT \cite{reimers2019sbert} is widely used for scalable similarity computation. Summarization work covers elicitation compression, contract obligation extraction, and code-to-requirement traceability \cite{spijkman2023, contractsumm2023, reposummary2025}. MARE \cite{jin2024mare} shows end-to-end RE automation, but targets modeling rather than test specification generation.

Despite these advances, a significant gap remains in the literature. Existing clustering-based approaches in SRE focus on organizing or analyzing requirements but do not exploit cluster topology to drive automated test generation at multiple levels of granularity. Similarly, LLM-based test generation approaches typically process requirements in isolation, discarding the inter-requirement context needed for integration testing. No prior work, to our knowledge, uses the semantic structure discovered through density-based clustering to automatically produce both individual requirement verification tests and cluster-level integration tests that verify cross-requirement feature behavior.

This paper addresses this gap by presenting a Cluster-then-Summarize pipeline specifically designed for large-scale automotive test specification generation under ASPICE SWE.6 constraints. We hypothesize that injecting cluster-derived context into LLM test generation (H1) increases integration coverage, (H2) improves grounding, and (H3) preserves summarization fidelity; Experiments~2--5 test these claims. Our contributions are as follows:

\begin{enumerate}
    \item A UMAP+HDBSCAN clustering pipeline with automatic \texttt{min\_cluster\_size} selection using normalized Silhouette and Calinski--Harabasz criteria.
    \item A multi-level map-reduce summarization method that preserves quantitative thresholds and safety levels.
    \item A dual-level test generation method for both individual requirements and cluster-level integration behavior, with nearby-cluster context injection.
    \item An empirical automotive evaluation showing improved integration coverage, lower redundancy, and stronger summarization fidelity versus baselines.
\end{enumerate}
\section{Related Work}

This section reviews four areas relevant to our pipeline: requirements clustering, requirements summarization, combined clustering-summarization pipelines, and LLM-based test generation.

\subsection{Requirements Clustering}
Clustering is widely used in requirements engineering to structure large requirement repositories. Early Bag-of-Words/TF-IDF approaches miss contextual semantics \cite{rosadadacruz2025ml}; dense embeddings, especially SBERT \cite{reimers2019sbert}, improve quality and enable efficient cosine-similarity search.
Multiple clustering paradigms have been applied to requirements data. K-Means has been used for test case minimization, demonstrating up to 75\% reduction in regression testing suites \cite{sharma2025kmeans}, though it requires specifying the number of clusters a priori \cite{clusteringtaxonomy2023}. Density-based algorithms DBSCAN \cite{ester1996dbscan} and HDBSCAN \cite{campello2013hdbscan} address this by automatically discovering cluster counts and handling varying densities. Hierarchical clustering has been applied to multi-document summarization and taxonomy generation \cite{hierarchicalsumm2014}, while graph-based overlapping clustering has been used for use case identification \cite{graphclustering2023}.
Beyond grouping, clustering has been applied to specific SRE tasks: transformer-based embeddings to detect semantic conflicts \cite{conflictdetection2022}, lexical shifts to infer document structure \cite{clusterstructure2014}, and external knowledge synthesis to automatically label clusters \cite{clusterlabeling2019}.
While these approaches demonstrate the value of clustering for organizing and analyzing requirements, none exploit the cluster topology itself to drive downstream test generation at multiple levels of granularity.
\subsection{Requirements Summarization}

SRE summarization includes extractive and abstractive methods \cite{abstractive2024survey}. Extractive methods preserve source wording but often reduce coherence \cite{rosadadacruz2025ml}, while abstractive Transformer models such as BART \cite{lewis2020bart} and GPT produce cleaner paraphrases but can hallucinate unsupported content \cite{abstractive2024survey}.

Domain-specific approaches include REConSum for summarizing elicitation conversations \cite{spijkman2023}, contract summarization for extracting obligations \cite{contractsumm2023}, and RepoSummary for restoring traceability in legacy codebases \cite{reposummary2025}. For large corpora, hierarchical summarization clusters documents and summarizes each group independently \cite{hierarchicalsumm2014,chatdialogue2021}. Our work extends this by using summaries as contextual inputs for downstream test generation.

\subsection{Combined Clustering-Summarization Pipelines}

Combining clustering and summarization is state of the art for large requirements corpora \cite{rosadadacruz2025ml}. A common pipeline embeds documents (e.g., SBERT/FastText), clusters them (K-Means/DBSCAN/HDBSCAN), then summarizes each cluster \cite{chatdialogue2021}. The result is a structured hierarchy where clusters capture high-level features \cite{hierarchicalsumm2014}.

At the systems level, the MARE framework \cite{jin2024mare} represents the most comprehensive end-to-end automation of requirements engineering, using five specialized LLM agents to handle elicitation, modeling, verification, and specification. Evaluated on public benchmarks including PURE \cite{pure2017}, MARE outperformed baselines by 15.4\% in modeling correctness. However, MARE targets requirements modeling and specification rather than test generation, and does not incorporate clustering as an architectural mechanism for organizing requirements at scale.

Other integrated approaches include the work of Siavashi et al. \cite{frontiers2025llmre}, who survey LLM applications across the full SRE lifecycle and identify test generation as an underexplored downstream task. Similarly, Zhao et al. \cite{nlp4re2020mapping} map NLP techniques to RE activities and note the absence of unified pipelines that bridge requirement analysis with verification artifact generation. Our pipeline fills this gap by treating clustering not as an analytical endpoint but as the structural scaffold for generating test specifications at two levels of granularity.

\subsection{LLM-Based Test Generation}

LLM-based test generation has grown rapidly, including prompt-based generation \cite{masuda2025ghl}, multi-step retrieval-generation pipelines \cite{adabala2025multistep}, and domain-specific fine-tuning \cite{chow2026automotive}. End-to-end automation has also been explored \cite{wang2025automotive}, along with structured prompting for control logic test cases \cite{koziolek2024control}.

However, these approaches process requirements individually, mapping each requirement to test cases in isolation. This lack of inter-requirement context prevents generation of integration tests that capture feature-level interactions. Our work addresses this limitation by clustering requirements and injecting cluster context into LLM calls, enabling both individual and cluster-level test generation.

\begin{figure*}[t]
\centering
\definecolor{cBlue}{HTML}{2980B9}
\definecolor{cTeal}{HTML}{1ABC9C}
\definecolor{cGreen}{HTML}{27AE60}
\definecolor{cOrange}{HTML}{E67E22}
\definecolor{cRed}{HTML}{C0392B}
\definecolor{cPurple}{HTML}{8E44AD}
\definecolor{cCyan}{HTML}{3498DB}
\definecolor{fbg}{HTML}{F7F9FA}
\definecolor{gry}{HTML}{7F8C8D}
\resizebox{0.95\textwidth}{!}{%
\begin{tikzpicture}[
    >={Stealth[length=4pt]},
    every node/.style={font=\footnotesize},
    arr/.style={->, thick, gry!80},
    darr/.style={->, thick, cCyan!80, dashed},
]

\newcommand{\stagebox}[6]{%
  \node[draw=#4, fill=fbg, rounded corners=2pt,
        minimum width=2.2cm, minimum height=0.9cm,
        line width=0.7pt, align=center, font=\footnotesize\bfseries]
        (#1) at (#2,#3) {#6};
  \draw[#4, line width=2.5pt, rounded corners=2pt]
        ([xshift=0.4pt]#1.north west) -- ([xshift=0.4pt]#1.south west);
  \node[circle, fill=#4, text=white, font=\tiny\bfseries,
        inner sep=0pt, minimum size=12pt]
        at ([shift={(-0.12cm,0.12cm)}]#1.north west) {#5};
}

\node[draw=gry!50, dashed, rounded corners=2pt, fill=white,
      minimum width=1.6cm, minimum height=0.8cm,
      align=center, font=\footnotesize]
      (inp) at (0, 0) {DOCX\\Reqs};

\stagebox{s1}{3.0}{0}{cBlue}{1}{Embedding}
\node[font=\tiny, text=gry, above=0.12cm of s1]
      {$\mathcal{E}(d_i)\!\in\!\mathbb{R}^{384}$};

\stagebox{s2}{6.0}{0}{cTeal}{2}{HDBSCAN\\Clustering}
\node[font=\tiny, text=gry, above=0.12cm of s2, xshift=0.2cm]
      {$\mu^*\!=\!\arg\max_{\mu\in\mathcal{M}} Q(\mu)$};

\foreach \yo/\col in {0.28/cBlue, 0.09/cTeal, -0.09/cGreen, -0.28/cOrange}{
    \fill[\col!30] (7.55, \yo) ellipse (0.2cm and 0.09cm);
    \foreach \a/\b in {-0.06/0.02, 0.05/-0.015, 0.0/0.03}{
        \fill[\col!80] ({7.55+\a},{\yo+\b}) circle (0.015cm);
    }
}

\stagebox{s3}{9.0}{0}{cGreen}{3}{Map-Reduce\\Summarize}
\node[font=\tiny, text=gry, above=0.12cm of s3]
      {$B\!=\!10,\;M\!=\!3$};

\stagebox{s4}{12.5}{1.2}{cOrange}{4}{Individual\\Tests}
\node[font=\tiny, text=gry, above=0.12cm of s4]
      {per requirement};

\stagebox{s5}{12.5}{-1.2}{cRed}{5}{Integration\\Tests}
\node[font=\tiny, text=gry, left=0.15cm of s5.south west, anchor=north east]
      {per cluster, $\geq\!2$ reqs};

\node[draw=cCyan, fill=cCyan!6, rounded corners=2pt,
      minimum width=1.6cm, minimum height=0.55cm,
      font=\tiny\bfseries, line width=0.6pt, align=center]
      (rag) at (12.5, -2.8) {RAG Store};
\node[font=\tiny, text=gry, below=0.06cm of rag]
      {ASPICE $\cdot$ ISO\,26262};

\node[draw=cPurple, fill=cPurple!5, rounded corners=2pt,
      minimum width=2.0cm, minimum height=1.6cm,
      line width=0.7pt, align=center, font=\footnotesize\bfseries]
      (out) at (15.8, 0) {$\mathcal{I}\!+\!\mathbf{T}^{\mathcal{I}}$\\[6pt]$\mathcal{G}\!+\!\mathbf{T}^{\mathcal{G}}$};
\draw[cPurple, line width=2.5pt, rounded corners=2pt]
      ([xshift=0.4pt]out.north west) -- ([xshift=0.4pt]out.south west);
\node[circle, fill=cPurple, text=white, font=\tiny\bfseries,
      inner sep=0pt, minimum size=12pt]
      at ([shift={(-0.12cm,0.12cm)}]out.north west) {$\Sigma$};

\draw[arr] (inp.east) -- (s1.west);
\draw[arr] (s1.east) -- (s2.west);
\draw[arr] (s2.east) -- ++(0.3,0);
\draw[arr] (7.85,0) -- (s3.west);
\draw[arr] (s3.east) -- ++(0.7,0) coordinate(split);
\draw[arr, rounded corners=4pt] (split) |- (s4.west);
\draw[arr, rounded corners=4pt] (split) |- (s5.west);
\draw[arr, rounded corners=4pt] (s4.east) -| (out.north);
\draw[arr, rounded corners=4pt] (s5.east) -| (out.south);

\draw[darr] (rag.north) -- (s5.south);
\draw[darr, rounded corners=2pt] ([xshift=0.8cm]rag.north) -- ++
(0,0.55) -- ++(0.55,0) |- ([yshift=-0.3cm]s4.east);

\node[font=\tiny, text=gry, anchor=west] at (0, -2.8) {
  $\mathcal{I}$: individual tests \;\;
  $\mathcal{G}$: integration tests \;\;
  $\mathbf{T}$: traceability \;\;
  $\sigma_j$: cluster summary \;\;
  \textcolor{cCyan}{- - -}: RAG context
};

\end{tikzpicture}%
}%
\caption{Architecture of the Cluster-then-Summarize pipeline. Requirements are embedded (Stage~1), grouped via HDBSCAN (Stage~2), and Map-Reduce summarization (Stage~3). The cluster topology drives two parallel paths: per-requirement individual tests (Stage~4) and per-cluster integration tests (Stage~5), both augmented by RAG-retrieved standards. Output comprises individual test specifications $\mathcal{I}$ with traceability $\mathbf{T}^{\mathcal{I}}$ and integration test specifications $\mathcal{G}$ with traceability $\mathbf{T}^{\mathcal{G}}$.}
\label{fig:pipeline}
\end{figure*}
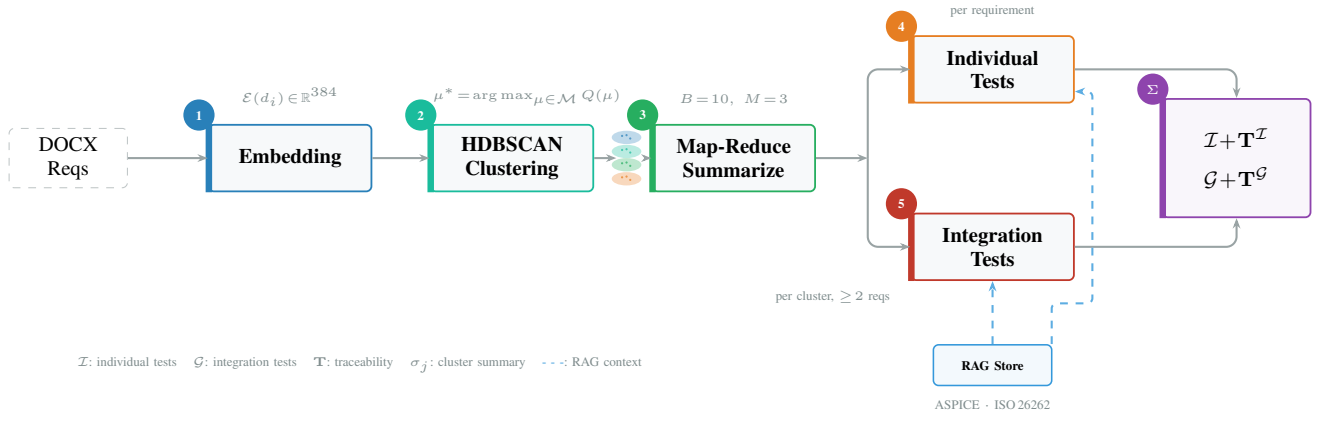
\section{Proposed Approach}

This section presents the Cluster-then-Summarize pipeline for dual-level test specification generation (Fig.~\ref{fig:pipeline}). The pipeline has five stages: (1) embedding and vector storage, (2) density-based clustering, (3) map-reduce cluster summarization, (4) cluster-aware semantic understanding and individual test generation, and (5) cluster-level integration test generation. Stages~1--3 process the full requirement set; Stages~4--5 run per requirement and per cluster respectively.


\subsection{Definitions and Notation}

Let $\mathcal{R} = \{r_1, r_2, \ldots, r_N\}$ denote the set of $N$ software requirements extracted from a project's specification documents. Each requirement $r_i$ consists of an identifier $\text{id}_i$, a natural language description $d_i$, metadata fields (safety level, category, domain), and a verification method. Let $\mathcal{E}: \mathcal{D} \rightarrow \mathbb{R}^m$ be a sentence embedding function that maps a textual description to an $m$-dimensional vector. We use \texttt{all-MiniLM-L6-v2}, a Sentence Transformer model, yielding $m = 384$.

\subsection{Stage 1: Embedding and Vector Storage}

Each requirement description $d_i$ is embedded independently:
\begin{equation}
    \mathbf{e}_i = \mathcal{E}(d_i), \quad \mathbf{e}_i \in \mathbb{R}^{384}
\end{equation}

Critically, we embed only the description field $d_i$, excluding the requirement identifier and metadata. This prevents requirements with adjacent IDs (e.g., \texttt{REQ-SW-042}, \texttt{REQ-SW-043}) from clustering based on lexical ID similarity rather than semantic content. All embeddings are stored in a ChromaDB vector store alongside three categories of supplementary documents: embedded chunks from automotive standards (ASPICE SWE.6, ISO~26262), contextual paragraphs from the requirements document, and narrative sections. These serve the RAG retrieval mechanism described in Section~\ref{sec:rag}.

\subsection{Stage 2: UMAP Dimensionality Reduction and HDBSCAN Clustering}

We apply a two-phase strategy to cluster the embedding matrix $\mathbf{E} = [\mathbf{e}_1, \ldots, \mathbf{e}_N]^\top$: first reducing dimensionality with UMAP \cite{mcinnes2018umap} to counteract the concentration-of-measure effect inherent in high-dimensional cosine spaces, then applying HDBSCAN \cite{campello2013hdbscan}, which discovers cluster counts automatically, accommodates clusters of varying density, and requires no global density threshold.

\subsubsection{UMAP Dimensionality Reduction}

We reduce each $384$-dimensional embedding $\mathbf{e}_i$ to a compact $d$-dimensional representation before clustering:
\begin{equation}
    \tilde{\mathbf{e}}_i = \text{UMAP}(\mathbf{e}_i), \quad \tilde{\mathbf{e}}_i \in \mathbb{R}^{d}
\end{equation}
UMAP is configured with $d = 15$, $n_{\text{neighbors}} = 15$, $\text{min\_dist} = 0.0$, and cosine metric on the input space. Setting $\text{min\_dist} = 0.0$ maximises local cluster compactness in the reduced space, which is optimal for the subsequent density estimation performed by HDBSCAN. The target dimension is clamped to $d = \min(15,\, m{-}1,\, N{-}2)$ to remain valid for any dataset size.

\subsubsection{Automatic \texttt{min\_cluster\_size} Selection}

HDBSCAN's primary parameter \texttt{min\_cluster\_size} ($\mu$) controls the minimum number of points required to form a cluster; groups smaller than $\mu$ are labelled as noise. Rather than fixing $\mu$ manually, we select it adaptively using a quality criterion that combines normalized Silhouette and normalized Calinski--Harabasz scores over a candidate set constructed from fixed seeds and dataset-size-relative fractions:
\begin{equation}
    \mathcal{M} = \left(\{2,3,5\} \cup \left\{\max\!\left(5,\left\lfloor\tfrac{N}{20}\right\rfloor\right),\, \max\!\left(5,\left\lfloor\tfrac{N}{10}\right\rfloor\right)\right\}\right) \cap \left[2,\left\lfloor\tfrac{N}{3}\right\rfloor\right]
\end{equation}
For each candidate $\mu \in \mathcal{M}$, a trial HDBSCAN run is performed on $\tilde{\mathbf{E}}$. Trial noise points are temporarily assigned to their nearest cluster centroid (see below) before metric computation. Two internal quality metrics are then evaluated: the Silhouette Score $\bar{s}(\mu)$ and the Calinski--Harabasz index $h(\mu)$. Both are min-max normalised across all candidates,
\begin{equation}
    \hat{s}(\mu) = \frac{\bar{s}(\mu) - \min_{\mathcal{M}} \bar{s}}{\max_{\mathcal{M}} \bar{s} - \min_{\mathcal{M}} \bar{s}}, \qquad
    \hat{h}(\mu) = \frac{h(\mu) - \min_{\mathcal{M}} h}{\max_{\mathcal{M}} h - \min_{\mathcal{M}} h}
\end{equation}
and combined into the selection criterion:
\begin{equation}
    Q(\mu) = 0.7\,\hat{s}(\mu) + 0.3\,\hat{h}(\mu)
\end{equation}
The optimal parameter is selected as:
\begin{equation}
    \mu^* = \arg\max_{\mu \in \mathcal{M}}\, Q(\mu)
\end{equation}
If no candidate yields a valid multi-cluster partition, $\mu^*$ falls back to $5$.

\subsubsection{HDBSCAN Clustering and Noise Reassignment}

HDBSCAN is applied to $\tilde{\mathbf{E}}$ using Euclidean distance and the selected $\mu^*$, assigning each requirement to a cluster or labelling it noise:
\begin{equation}
    \ell_i = \text{HDBSCAN}_{\mu^*}(\tilde{\mathbf{e}}_i), \qquad \ell_i \in \{-1, 0, 1, \ldots, K{-}1\}
\end{equation}
where $\ell_i = -1$ denotes noise and $K$ is the number of clusters discovered.

Rather than discarding noise points or promoting them to singleton clusters, each noise point is reassigned to the nearest existing cluster by cosine similarity to cluster centroids computed in the UMAP-reduced space. For cluster $C_j$, its centroid in the reduced space is:
\begin{equation}
    \tilde{\boldsymbol{\mu}}_j = \frac{1}{|C_j|} \sum_{r_i \in C_j} \tilde{\mathbf{e}}_i
\end{equation}
Each noise point $r_n$ ($\ell_n = -1$) is then assigned to cluster $C_{j^*}$ where:
\begin{equation}
    j^* = \arg\max_j\; \frac{\tilde{\boldsymbol{\mu}}_j \cdot \tilde{\mathbf{e}}_n}{\|\tilde{\boldsymbol{\mu}}_j\|\,\|\tilde{\mathbf{e}}_n\|}
\end{equation}
This yields a complete partition $\mathcal{C} = \{C_1, \ldots, C_K\}$ satisfying $\bigcup_{C \in \mathcal{C}} C = \mathcal{R}$, preserving full requirement coverage without artificially inflating the cluster count with singletons.

\subsubsection{Nearby-Cluster Computation}

For each cluster $C_j$, we compute its centroid in the original high-dimensional embedding space to capture full semantic content:
\begin{equation}
    \boldsymbol{\mu}_j = \frac{1}{|C_j|} \sum_{r_i \in C_j} \mathbf{e}_i
\end{equation}

The $\kappa = 3$ nearest clusters are identified by centroid-to-centroid cosine distance:
\begin{equation}
    \mathcal{N}(C_j) = \underset{C_l \in \mathcal{C} \setminus \{C_j\}}{\arg\text{top-}\kappa} \; \text{dist}_{\cos}(\boldsymbol{\mu}_j, \boldsymbol{\mu}_l)
    \label{eq:nearby}
\end{equation}

These nearby clusters provide bounded cross-feature context during semantic understanding (Stage~4), enabling the LLM to reason about inter-feature dependencies without loading the entire requirement corpus.

\subsection{Stage 3: Map-Reduce Cluster Summarization}

Each cluster $C_j$ is distilled into a concise, domain-conformant description $\sigma_j$ that captures the collective functional intent of its member requirements. Because direct concatenation of large clusters can exceed LLM context limits, we apply a multi-level map-reduce strategy parameterized by requirement batch size $B$ and summary merge factor $M$. In the implementation used in this work, $B=10$ and $M=3$.

\subsubsection{Formal Definition}

Let $\text{LLM}_r(S)$ denote summarization of a set of requirements $S$ into one description, and let $\text{LLM}_s(S)$ denote synthesis of a set of intermediate summaries $S$ into one summary.

For cluster $C_j$ with $n_j = |C_j|$, define the requirement partition into $K_j = \left\lceil n_j / B \right\rceil$ batches:
\begin{equation}
    \mathcal{P}_B(C_j) = \{R_{j,1}, \ldots, R_{j,K_j}\},
    \qquad K_j = \left\lceil \frac{n_j}{B} \right\rceil
\end{equation}

The map step generates one summary per requirement batch:
\begin{equation}
    s_{j,k} = \text{LLM}_r(R_{j,k}),
    \qquad k = 1, \ldots, K_j
\end{equation}

Define the level-0 summary set as:
\begin{equation}
    S_j^{(0)} = \{s_{j,1}, \ldots, s_{j,K_j}\}
\end{equation}

The reduce step then repeatedly merges summaries in groups of at most $M$:
\begin{equation}
    S_j^{(\ell+1)} = \left\{\text{LLM}_s(G)\;\middle|\; G \in \mathcal{P}_M\left(S_j^{(\ell)}\right)\right\},
    \qquad \ell = 0,1,2,\ldots
\end{equation}

Reduction stops at the first level $L_j$ such that $\left|S_j^{(L_j)}\right| = 1$, and the final cluster description is:
\begin{equation}
    \sigma_j = \text{the unique element of } S_j^{(L_j)}
\end{equation}

When $n_j \leq B$, the method degenerates to a single call $\text{LLM}_r(C_j)$, which is the base case of the same map-reduce algorithm. With $B=10$ and $M=3$, each cluster's summarization requires $\mathcal{O}(\frac{3n}{20})$ LLM calls while preserving quantitative fidelity and inheriting the maximum ASIL safety level across cluster members.

\subsection{Stage 4: Cluster-Aware Semantic Understanding and Individual Test Generation}

Each requirement $r_i \in C_j$ is analyzed by the LLM with a context window that incorporates information from multiple sources. This is the key mechanism through which clustering improves per-requirement test quality.

\subsubsection{Context Assembly}
\label{sec:rag}

For requirement $r_i$ belonging to cluster $C_j$, the assembled context $\Gamma(r_i)$ consists of six components:
\begin{equation}
    \Gamma(r_i) = \{r_i\} \cup C_j \cup \{\sigma_j\} \cup \mathcal{N}_r(C_j) \cup \text{RAG}_{\text{std}} \cup \text{RAG}_{\text{ctx}}(r_i)
\end{equation}
where $\sigma_j$ is the cluster description from Stage~3, $\mathcal{N}_r(C_j)$ denotes requirements sampled from the nearby clusters $\mathcal{N}(C_j)$ defined in Eq.~\ref{eq:nearby}, $\text{RAG}_{\text{std}}$ represents cached standard guideline chunks, and $\text{RAG}_{\text{ctx}}(r_i)$ represents requirement-specific contextual paragraphs retrieved via cosine similarity.

To bound context length, when $|C_j \cup \mathcal{N}_r(C_j)| > \tau$ (with $\tau = 20$), a random sample of $\tau$ requirements is drawn. RAG retrieval applies a minimum cosine similarity threshold $\theta = 0.6$; chunks below this threshold are discarded.

\subsubsection{Semantic Analysis Output}

The LLM produces a structured analysis for each requirement containing: a natural language interpretation of the requirement's meaning, a set of key behaviors $\mathcal{B}(r_i) = \{b_1, \ldots, b_p\}$, testable conditions $\mathcal{T}(r_i) = \{t_1, \ldots, t_q\}$, and a set of test intents with estimated count $n_i \in [1, 5]$.

\subsubsection{Individual Test Specification Generation}

For each requirement $r_i$, the pipeline generates $n_i$ test specifications. Each test specification $\text{TS}_{i,k}$ ($k = 1, \ldots, n_i$) is a structured record:
\begin{equation}
    \text{TS}_{i,k} = \langle \text{id}, \text{desc}, \text{pre}, \text{in}, \text{exp}, \text{pass}, \text{post}, \text{proc} \rangle
\end{equation}
containing a unique identifier, description, preconditions derived from $\mathcal{T}(r_i)$, input parameters, expected results derived from $\mathcal{B}(r_i)$, measurable pass criteria, postconditions, and a step-by-step execution procedure. The total individual test set is:
\begin{equation}
    \mathcal{I} = \bigcup_{i=1}^{N} \{\text{TS}_{i,1}, \ldots, \text{TS}_{i,n_i}\}
\end{equation}

\subsection{Stage 5: Cluster-Level Integration Test Generation}

This stage is the primary novel contribution of the pipeline. For each cluster $C_j$ with $|C_j| \geq 2$, the pipeline generates feature-level integration tests that verify cross-requirement behavior.

\subsubsection{Context Preparation}

For each cluster $C_j$, the context is assembled as:
\begin{equation}
    \Phi(C_j) = \text{extract}(C_j) \cup \{\sigma_j\} \cup \text{RAG}_{\text{std}} \cup \text{RAG}_{\text{ctx}}(C_j)
\end{equation}
where $\text{extract}(C_j)$ reduces each requirement to only its identifier and description.

\subsubsection{Integration Test Constraints}

The LLM is constrained to generate only integration-level tests. Each cluster test specification $\text{IT}_{j,k}$ must satisfy:
\begin{equation}
    |\text{traced}(\text{IT}_{j,k})| \geq 2
\end{equation}
where $\text{traced}(\text{IT}_{j,k}) \subseteq C_j$ is the set of requirements verified by the test. The total integration test set is:
\begin{equation}
    \mathcal{G} = \bigcup_{j : |C_j| \geq 2} \{\text{IT}_{j,1}, \ldots, \text{IT}_{j,m_j}\}
\end{equation}

\subsubsection{Traceability}

The pipeline produces two traceability matrices. The individual traceability matrix $\mathbf{T}^{\mathcal{I}} \in \{0,1\}^{N \times |\mathcal{I}|}$ maps requirements to individual tests:
\begin{equation}
    T^{\mathcal{I}}_{i,k} = 
    \begin{cases}
        1 & \text{if } \text{TS}_k \text{ traces to } r_i \\
        0 & \text{otherwise}
    \end{cases}
\end{equation}

The cluster traceability matrix $\mathbf{T}^{\mathcal{G}} \in \{0,1\}^{N \times |\mathcal{G}|}$ maps requirements to integration tests:
\begin{equation}
    T^{\mathcal{G}}_{i,k} = 
    \begin{cases}
        1 & \text{if } r_i \in \text{traced}(\text{IT}_k) \\
        0 & \text{otherwise}
    \end{cases}
\end{equation}

Together, these ensure that every requirement $r_i$ is traced by at least one individual test ($\sum_k T^{\mathcal{I}}_{i,k} \geq 1$), and requirements in multi-member clusters are additionally covered by integration tests ($\sum_k T^{\mathcal{G}}_{i,k} \geq 1$ for $r_i \in C_j$ with $|C_j| \geq 2$).

\section{Experiments}

We use seven datasets ($D_1$--$D_7$) in this section. Unless stated otherwise, experiments are run on all seven. Experiment~1 is the only exception and uses $D_1$--$D_5$ because it analyzes clustering behavior versus dataset size.

\subsection{Experiment 1: Clustering Quality Across Scales}

\subsubsection{Objective}

We evaluate clustering on five datasets ($D_1$--$D_5$) of increasing size. The aim is to compare HDBSCAN, K-Means, and DBSCAN in quality and scalability.

\subsubsection{Datasets}

Experiment~1 uses five datasets: $D_1$ (33 requirements), $D_2$ (57), $D_3$ (171), $D_4$ (219), and $D_5$ (521).

All datasets contain a mix of functional and non-functional requirements. Each requirement was embedded using \texttt{all-MiniLM-L6-v2} (384 dimensions) with description-only embeddings, as described in Stage~1 of the methodology.

\subsubsection{Experimental Setup}

All methods use UMAP-reduced embeddings ($d=15$, cosine input metric) to ensure fair comparison and reduce high-dimensional distance effects.

We evaluate three algorithms. \textbf{K-Means}: elbow search over $K \in \{1,\ldots,20\}$. \textbf{DBSCAN}: $\varepsilon$ selected by k-distance percentile search $p \in \{25,50,75,90\}$ with $\text{min\_samples}=2$. \textbf{HDBSCAN}: automatic \texttt{min\_cluster\_size} selection using $Q(\mu)=0.7\,\hat{s}(\mu)+0.3\,\hat{h}(\mu)$ over $\mathcal{M}$ (Section~3.2).

Noise points produced by DBSCAN are reassigned to the nearest cluster by cosine similarity before metric computation. HDBSCAN noise points are reassigned in the same manner, as described in Section~3.2.

All experiments were run with a fixed random seed (42) for reproducibility.

\subsubsection{Evaluation Metrics}

We report four metrics: \textbf{Silhouette} ($\bar{s}$), \textbf{Calinski--Harabasz (CH) index}, \textbf{number of clusters} ($K$), and \textbf{noise ratio} ($\rho$; zero for K-Means, before reassignment for density-based methods).

\subsubsection{Results}

Table~\ref{tab:results_all} reports all clustering metrics for the three algorithms across all five datasets. Best result per dataset per metric is \textbf{bold}.

\begin{table}[htbp]
\centering
\caption{Clustering results across dataset scales. $\rho$: fraction of requirements initially labelled noise; K-Means produces no noise by construction (N/A). Best result is \textbf{bold}.}
\label{tab:results_all}
\resizebox{\columnwidth}{!}{
\setlength{\tabcolsep}{6pt}
\begin{tabular}{llcccc}
\toprule
\textbf{Dataset} & \textbf{Algorithm} & \textbf{$K$} & \textbf{Silhouette} & \textbf{CH Index} & \textbf{$\rho$} \\
\midrule
\multirow{3}{*}{$D_1$ (33)}
  & K-Means & 2 & 0.5944 & 90.83 & N/A \\
  & DBSCAN  & 5  & 0.4349 & 55.16     & 0.1212 \\
  & HDBSCAN & 2  & \textbf{0.6631} & \textbf{136.07} & \textbf{0.0000} \\
\midrule
\multirow{3}{*}{$D_2$ (57)}
  & K-Means & 8 & 0.4247 & 48.80 & N/A \\
  & DBSCAN  & 9  & 0.4380 & 57.94     & \textbf{0.1053} \\
  & HDBSCAN & 8  & \textbf{0.4505} & \textbf{60.81}  & 0.1404 \\
\midrule
\multirow{3}{*}{$D_3$ (171)}
  & K-Means & 2 & 0.4309 & 174.81 & N/A \\
  & DBSCAN  & 31 & 0.3947 & 190.17    & 0.0994 \\
  & HDBSCAN & 5  & \textbf{0.4918} & \textbf{195.79} & \textbf{0.0351} \\
\midrule
\multirow{3}{*}{$D_4$ (219)}
  & K-Means & 8 & 0.4651 & 215.54 & N/A \\
  & DBSCAN  & 42 & 0.4712 & \textbf{350.64}    & \textbf{0.1005} \\
  & HDBSCAN & 5  & \textbf{0.4944} & 287.05  & 0.1826 \\
\midrule
\multirow{3}{*}{$D_5$ (521)}
  & K-Means & 20 & \textbf{0.9577} & 10245.04 & N/A \\
  & DBSCAN  & 62 & 0.6574 & \textbf{243125.81} & 0.0998 \\
  & HDBSCAN & 38 & 0.7967 & 230546.19 & \textbf{0.0058} \\
\bottomrule
\end{tabular}
}
\end{table}

\subsubsection{Analysis}

HDBSCAN achieves the highest Silhouette Score on four of five datasets and produces the lowest noise ratio $\rho$ on four of five datasets, indicating consistently compact and well-separated clusters with minimal outlier sensitivity. The K-Means outlier on $D_5$ (Silhouette = 0.9577, $K=20$) reflects the elbow search hitting its upper bound, yielding many small, artificially tight clusters rather than semantically meaningful groupings. DBSCAN's superiority in CH Index on $D_4$--$D_5$ is accompanied by extreme over-fragmentation ($K = 42$ and $K = 62$ respectively), making its clusters impractical for downstream test specification generation. HDBSCAN produces stable, interpretable cluster counts across all scales with no manual parameter tuning, making it the preferred choice for the pipeline.

\subsection{Experiment 2: Summarization Strategy Comparison}
This experiment compares three LLM summarization strategies (Single-Pass, Map-Reduce with $B=10$, $M=3$, and Recursive) on all datasets ($D_1$--$D_7$). Each summary is evaluated against human references using ROUGE-L, BERTScore, and expert ratings from three ASPICE-experienced engineers on Completeness, Quantitative Preservation, and Conciseness (1--5 Likert, majority vote; ICC = 0.81). The full results are reported in Table~\ref{tab:exp2_results}.

\begin{table}[htbp]
\centering
\caption{Summarisation strategy comparison. Results are aggregated across datasets; expert scores are averaged per cluster (scale 1--5). Best per metric \textbf{bold}.}
\label{tab:exp2_results}
\resizebox{\columnwidth}{!}{
\begin{tabular}{lccccc}
\toprule
\textbf{Strategy} & \textbf{ROUGE-L} & \textbf{BERTScore} & \textbf{Completeness} & \textbf{Quant. Pres.} & \textbf{Conciseness} \\
\midrule
Single-Pass & 0.2934          & 0.8776             & 3.19                  & 3.38                  & \textbf{4.19}        \\
Map-Reduce  & \textbf{0.3793} & \textbf{0.8908}    & \textbf{4.30}         & \textbf{4.34}         & 3.08                 \\
Recursive   & 0.3439          & 0.8875             & 3.98                  & 3.82                  & 3.23                 \\
\bottomrule
\end{tabular}
}
\end{table}

\subsubsection{Analysis}
Table~\ref{tab:exp2_results} shows aggregated performance across datasets.
Map-reduce is the strongest strategy: it achieves the highest ROUGE-L (0.3793), the highest BERTScore (0.8908), the highest completeness (4.30), and the highest quantitative preservation (4.34). Relative to single-pass, map-reduce improves completeness by 1.11 points and quantitative preservation by 0.96 points, while also improving both automatic metrics.
Single-pass remains the most concise strategy (4.19), but this conciseness comes with lower semantic and content fidelity. The recursive baseline is more balanced than single-pass on fidelity-oriented metrics, yet remains below map-reduce on ROUGE-L, BERTScore, completeness, and quantitative preservation.
Based on these results, the pipeline adopts map-reduce as the default summarisation strategy for all clusters. For small clusters, the method naturally degenerates to a single map-level call without additional reduce levels.

While LLM-generated summaries risk omitting content or introducing inaccuracies, the experimental results empirically address this concern: the best strategy achieves a BERTScore of 0.8908 with expert completeness and quantitative preservation ratings of 4.30 and 4.34 out of 5, demonstrating strong fidelity to the original source documents.

\subsection{Experiment 3: Test Generation Quality with Cluster Context}

This experiment isolates cluster context effects on LLM test generation across all datasets. For each requirement, tests are generated twice: Condition A (baseline, requirement only) and Condition B (with cluster context $C_j$, summary $\sigma_j$, and nearby requirements).

Evaluation uses \textbf{Semantic Diversity} (pairwise cosine distance) of SBERT embeddings (\texttt{all-MiniLM-L6-v2}) , \textbf{METEOR} measures recall-oriented lexical alignment between each test and its linked requirement, macro-averaged per requirement. \textbf{Specificity} is a composite score, combining (i)~numeric density: the frequency of measurable values with physical units per token, (ii)~clause-level ratio of structured identifiers and measurable tokens, (iii)~MATTR: a length-normalized vocabulary richness measure that rewards lexical variety without penalizing longer texts, and (iv)~inverse proportion of generic and hedging terms.
\textbf{Overlap Error Rate} flags tests where the semantic margin
$\Delta = \text{sim}(t,\, r_{\text{true}}) -\max_{r \neq r_{\text{true}}} \text{sim}(t,\, r) < -0.02$, indicating a semantic boundary violation beyond the noise threshold.

\subsubsection{Results}

Table~\ref{tab:exp3_results} summarizes the performance of Conditions A and B. Here, $\bar{I}$ denotes the mean percentage improvement across the corpus.

\begin{table}[htbp]
\centering
\caption{Comparison of Condition A and Condition B. Best values per row are highlighted in bold.}
\label{tab:exp3_results}
\resizebox{\columnwidth}{!}{
\begin{tabular}{l|cc|cc|cc|cc}
\toprule
\textbf{Dataset} & \multicolumn{2}{c|}{\textbf{Semantic Diversity}} & \multicolumn{2}{c|}{\textbf{METEOR}} & \multicolumn{2}{c|}{\textbf{Specificity}} & \multicolumn{2}{c}{\textbf{Overlap Error (\%)}} \\
 & \textbf{A} & \textbf{B} & \textbf{A} & \textbf{B} & \textbf{A} & \textbf{B} & \textbf{A} & \textbf{B} \\
\midrule
$D_1$  & 0.696 & \textbf{0.721} & 0.297 & \textbf{0.343} & 0.292 & \textbf{0.366} & 22.33 & \textbf{17.02} \\
$D_2$  & 0.660 & \textbf{0.701} & 0.257 & \textbf{0.325} & 0.384 & \textbf{0.434} & 17.22 & \textbf{14.63} \\
$D_3$  & 0.648 & \textbf{0.693} & 0.285 & \textbf{0.329} & 0.602 & \textbf{0.695} & 28.63 & \textbf{21.79} \\
$D_4$  & 0.641 & \textbf{0.665} & 0.250 & \textbf{0.282} & 0.607 & \textbf{0.624} &  \textbf{24.00} & 25.82 \\
$D_5$  & 0.653 & \textbf{0.661} & 0.306 & \textbf{0.329} & \textbf{0.246} & 0.235 &  \textbf{27.60} &  27.98 \\
$D_6$  & 0.645 & \textbf{0.669} & 0.274 & \textbf{0.314} & \textbf{0.196} & 0.181 & 28.29 & \textbf{20.82} \\
$D_7$ & 0.661 & \textbf{0.687} & 0.292 & \textbf{0.325} & 0.225 & \textbf{0.263} & 22.83 & \textbf{17.72} \\
\midrule
\textbf{Mean ($\bar{I}$)} & \multicolumn{2}{c|}{+4.21\%} & \multicolumn{2}{c|}{+14.75\%} & \multicolumn{2}{c|}{+8.79\%} & \multicolumn{2}{c}{+14.65\%} \\
\bottomrule
\end{tabular}
}
\end{table}

\subsubsection{Analysis}

Providing cluster context consistently enhances the quality of LLM-generated tests across all metrics. Condition B enables the model to better recognize semantic boundaries, reducing overlap errors and improving specificity. Cluster context also promotes semantic diversity by encouraging the generation of distinct tests, while reducing generic, broad phrasing.

Slight reductions in specificity for $D_5$ and $D_6$ suggest that dense clusters can induce more general wording. Overall, cluster context remains essential for high-fidelity, precise, boundary-aware test generation.
\subsection{Experiment 4: Dual-Level Test Coverage Analysis}
\label{sec:experiment_integration}

This experiment evaluates whether cluster-level integration tests capture cross-requirement behavior missed by individual-only generation. We compare test counts between an individual-only baseline and the proposed dual-level pipeline across all datasets ($D_1$--$D_7$), additionally, we show percentage of requirements being mapped to cluster-level test specifications for every dataset.

\begin{table}[htbp]
\caption{Generated Test Specifications and Requirements Mapping}
\label{tab:test_counts}
\resizebox{\columnwidth}{!}{
\centering
\begin{tabular}{lccc}
\toprule
\textbf{Dataset} & \textbf{Baseline (Individual)} & \textbf{Dual-Level} & \textbf{Requirements Mapped to Cluster tests(\%) } \\
\midrule
$D_1$ & 88  & \textbf{107}  & 67\% \\
$D_2$ & 171 & \textbf{210}  & 85\% \\
$D_3$ & 462 & \textbf{609} & 78\% \\
$D_4$ & 484 & \textbf{582} & 66\% \\
$D_5$ & 510 & \textbf{587}  & 63\% \\
$D_6$ & 612 & \textbf{633}  & 20\% \\
$D_7$ & 1613 & \textbf{1766} & 41\% \\

\bottomrule
\end{tabular}
}
\end{table}

Table~\ref{tab:test_counts} shows test counts across all datasets. Review of the generated test specifications reveals broader coverage of complex behavioral interactions under the cluster-level approach. The following examples, drawn from a Surround-View Automated Parking System (SV-APA) requirements dataset (D3), illustrate integration scenarios captured exclusively by the proposed method.

The first sample test case verifies the integration between \texttt{REQ\_00071}, which requires the system to define proximity alert zones at graduated distances from the vehicle and issue escalating warnings as objects enter
successively closer zones, and \texttt{REQ\_00160}, which mandates that
collision avoidance take priority over parking maneuver completion in all situations. The cluster-level test case takes place when a parking maneuver
is in action, as it verifies the system capability of simultaneously producing the correct zone-level warnings and suppressing parking maneuver continuation
in favor of collision avoidance -- a compound behavior that no
individual-requirement test can exercise.

The second test case targets a distinct integration concern at the HMI layer, combining \texttt{REQ\_00167}, which constrains HMI response time to no more than 200\,ms from user input to feedback, with \texttt{REQ\_00085}, which requires the HMI to support display of text and voice guidance in multiple driver-selectable languages. Individually, each requirement is verifiable in a default system configuration. However, the integration test verifies that the 200\,ms responsiveness constraint is maintained uniformly across all supported languages -- a condition that only becomes observable when both requirements are active in the same test context.

Together, the quantitative coverage gains in Table~\ref{tab:test_counts} and the qualitative examples support H1. Basic LLM approaches that process requirements in isolation inherently fail to verify complex system interactions, creating a critical gap in ASPICE SWE.6 compliance.
\subsection{Experiment 5: Hallucination and Faithfulness in Test Generation}
This experiment evaluates grounding in generated individual and 
integration tests using Condition A (baseline, requirement only) and 
Condition B (cluster-aware with context). An automated detector flags 
four hallucination categories: \textbf{numeric} and \textbf{entity} 
grounding rely on hierarchical matching against linked requirements, 
cluster-level context, and dataset-wide references, combined with 
\textbf{SBERT} similarity thresholds and numeric tolerance checks. 
\textbf{Action-verb} hallucinations flag safety-critical control verbs 
absent from linked requirements via lemmatized matching, while 
\textbf{condition} hallucinations detect unsupported operational states 
via SBERT similarity. The \emph{flag rate} is the fraction of tests 
containing at least one hallucination, and \emph{faithfulness} is 
defined as $100 - \text{flag rate}$.

\begin{table}[htbp]
\centering
\caption{Summary of hallucination metrics using the grounding detector. Lower values indicate better grounding.}

\resizebox{\columnwidth}{!}{
\begin{tabular}{lccccc}
\toprule
\textbf{Dataset} & \textbf{Flag rate A} & \textbf{Flag rate B} & \textbf{Avg/test A} & \textbf{Avg/test B} & \textbf{Faithfulness (\%)} \\
\midrule
D1 & 25.09\% & 10.44\% & 0.317 & 0.185 & 89.56 \\
D2 & 51.46\% & 15.96\% & 0.660 & 0.512 & 84.04 \\
D3 & 28.23\% & 11.12\% & 0.321 & 0.144 & 88.88 \\
D4 & 11.93\% & 10.02\% & 0.149 & 0.131 & 89.98 \\
D5 & 17.19\% & 11.21\% & 0.179 & 0.154 & 88.79 \\
D6 & 10.52\% & 9.98\% & 0.105 & 0.101 & 90.02 \\
D7 & 13.18\% & 9.94\% & 0.149 & 0.135 & 90.06 \\
\midrule
\textbf{Average} &  & \textbf{+17.09\%} &  & \textbf{+13.50\%} & \textbf{89.59\%} \\
\bottomrule
\end{tabular}
}

\label{tab:hallucination_results}
\end{table}

To further analyze grounding reliability, Table~\ref{tab:cluster_faithfulness} reports detailed hallucination statistics for cluster-level generated tests.

\begin{table}[htbp]
\centering
\small
\setlength{\tabcolsep}{5pt}
\renewcommand{\arraystretch}{1.15}
\caption{Faithfulness evaluation on cluster-level test specifications.}
\label{tab:cluster_faithfulness}
\begin{tabular}{lccc}
\toprule
\textbf{Dataset} & \textbf{Flag rate} & \textbf{Avg items / test} & \textbf{Faithfulness} \\
\midrule
D1  & \textbf{10.53\%} & \textbf{0.105} & \textbf{89.47\%} \\
D2  & \textbf{10.26\%} & \textbf{0.103} & \textbf{89.74\%} \\
D3 & \textbf{10.39\%} & \textbf{0.117} & \textbf{89.61\%} \\
D4 & \textbf{4.76\%}  & \textbf{0.048} & \textbf{95.24\%} \\
D5 & \textbf{7.84\%}  & \textbf{0.078} & \textbf{92.16\%} \\

D6 & \textbf{7.06\%}  & \textbf{0.071} & \textbf{92.94\%} \\
D7 & \textbf{10.11\%}  & \textbf{0.101} & \textbf{89.89\%} \\

\midrule
\textbf{Average} 
& {\textbf{8.71\%}} 
& {\textbf{0.089}} 
& {\textbf{91.29\%}} \\
\bottomrule
\end{tabular}

\end{table}

\subsubsection{Analysis}

Table~\ref{tab:hallucination_results} summarizes hallucination metrics across all datasets.
The detector uses stricter grounding logic for numeric values, units, and contextual references, reducing false positives while keeping a conservative safety bias.

Table~\ref{tab:cluster_faithfulness} confirms reliable cluster-level generation: average faithfulness is \textbf{91.29\%}, with low hallucination intensity (0.0890 items/test). These results support cluster-aware grounding for safety-critical automotive test generation.
\section{Conclusion}

This paper presented a Cluster-then-Summarize pipeline for automotive test specification generation under ASPICE SWE.6 constraints. The approach combines SBERT embeddings, UMAP+HDBSCAN clustering with adaptive \texttt{min\_cluster\_size} selection, map-reduce summarization, and cluster-aware dual-level test generation (individual and integration tests). By using cluster topology and nearby-cluster context, the pipeline preserves requirement semantics while capturing cross-requirement interactions that isolated prompting often misses.

Evaluation across seven datasets (with Experiment~1 run on five datasets for clustering-scale analysis) shows consistent gains. HDBSCAN produced stable and interpretable clusters across scales, map-reduce summarization achieved the strongest fidelity metrics, and cluster context improved test generation quality on average (+4.21\% semantic diversity, +14.75\% METEOR, +8.79\% specificity, and +14.65\% overlap-error reduction). Dual-level generation increased coverage by adding integration scenarios beyond an individual-only baseline. Hallucination analysis also showed strong grounding, with 89.59\% overall faithfulness and 91.29\% faithfulness for cluster-level tests.

Overall, these results confirm that the proposed pipeline successfully meets the paper's primary goal of accelerating the generation of reliable, ASPICE SWE.6-compliant test specifications. By demonstrating that clustering is not only an organizational step, but an effective structural basis for LLM-driven generation, this approach provides a robust solution to the manual engineering bottleneck inherent in safety-critical domains. Future work will focus on expert-in-the-loop validation, adaptive context selection for very large clusters, and tighter integration with industrial ALM toolchains for end-to-end traceability and continuous updates.

\bibliographystyle{IEEEtran}
\bibliography{ref}

@inproceedings{campello2013hdbscan,
  author    = {Ricardo J. G. B. Campello and
               Davoud Moulavi and
               J{\"o}rg Sander},
  title     = {Density-Based Clustering Based on Hierarchical Density Estimates},
  booktitle = {Proceedings of the Pacific-Asia Conference on Knowledge Discovery and Data Mining (PAKDD)},
  series    = {Lecture Notes in Computer Science},
  volume    = {7819},
  pages     = {160--172},
  year      = {2013},
  publisher = {Springer},
  doi       = {10.1007/978-3-642-37456-2_14}
}

@article{mcinnes2018umap,
  title={UMAP: Uniform Manifold Approximation and Projection for Dimension Reduction},
  author={McInnes, Leland and Healy, John and Melville, James},
  journal={arXiv preprint arXiv:1802.03426},
  year={2018}
}

@article{necula2024slr,
  author    = {Necula, Sabina-Cristiana and Dumitriu, Florin and Greavu-{\c{S}}erban, Vasile},
  title     = {A Systematic Literature Review on Using Natural Language Processing in Software Requirements Engineering},
  journal   = {Electronics},
  volume    = {13},
  number    = {11},
  pages     = {2055},
  year      = {2024},
  publisher = {MDPI},
  doi       = {10.3390/electronics13112055}
}

@article{rosadadacruz2025ml,
  author    = {Rosado da Cruz, Ant{\'o}nio Miguel and Cruz, Estrela Ferreira},
  title     = {Machine Learning Techniques for Requirements Engineering: A Comprehensive Literature Review},
  journal   = {Software},
  volume    = {4},
  number    = {3},
  pages     = {14},
  year      = {2025},
  publisher = {MDPI},
  doi       = {10.3390/software4030014}
}

@inproceedings{spijkman2023,
  author    = {Spijkman, Tjerk and de Bondt, Xavier and Dalpiaz, Fabiano and Brinkkemper, Sjaak},
  title     = {Summarization of Elicitation Conversations to Locate Requirements-Relevant Information},
  booktitle = {International Working Conference on Requirements Engineering: Foundation for Software Quality (REFSQ)},
  series    = {Lecture Notes in Computer Science},
  volume    = {13975},
  pages     = {127--143},
  year      = {2023},
  publisher = {Springer},
  doi       = {10.1007/978-3-031-29786-1\_9}
}

@article{jin2024mare,
  author    = {Jin, Dongming and Jin, Zhi and Chen, Xiaohong and Wang, Chunhui},
  title     = {{MARE}: Multi-Agents Collaboration Framework for Requirements Engineering},
  journal   = {arXiv preprint arXiv:2405.03256},
  year      = {2024},
  doi       = {10.48550/arXiv.2405.03256}
}

@inproceedings{reimers2019sbert,
  author    = {Reimers, Nils and Gurevych, Iryna},
  title     = {Sentence-{BERT}: Sentence Embeddings using Siamese {BERT}-Networks},
  booktitle = {Proceedings of the 2019 Conference on Empirical Methods in Natural Language Processing and the 9th International Joint Conference on Natural Language Processing (EMNLP-IJCNLP)},
  pages     = {3982--3992},
  year      = {2019},
  publisher = {Association for Computational Linguistics},
  address   = {Hong Kong, China},
  doi       = {10.18653/v1/D19-1410}
}

@article{llm4re2025,
  author    = {Nascimento, Nathalia and Santos, Amadeu and Lucena, Carlos},
  title     = {Large Language Models ({LLMs}) for Requirements Engineering ({RE}): A Systematic Literature Review},
  journal   = {arXiv preprint arXiv:2509.11446},
  year      = {2025}
}

@inproceedings{nlp4re2020mapping,
  author    = {Zhao, Liping and Alhoshan, Waad and Ferrari, Alessio and Letsholo, Keletso J. and Ajagbe, Muideen A. and Batista-Navarro, Riza Theresa and Sherwood, Michael},
  title     = {Natural Language Processing for Requirements Engineering: A Systematic Mapping Study},
  journal   = {ACM Computing Surveys},
  volume    = {54},
  number    = {3},
  pages     = {1--41},
  year      = {2022},
  publisher = {ACM},
  doi       = {10.1145/3444689}
}

@article{abstractive2024survey,
  author    = {Abdelaal, Hadi and Kabir, Moinul and Phan, Khang Nhut and Fuad, Shafin Rahman},
  title     = {Abstractive Text Summarization: State of the Art, Challenges, and Improvements},
  journal   = {arXiv preprint arXiv:2409.02413},
  year      = {2024}
}

@mastersthesis{chatdialogue2021,
  author    = {Jonsson, Tobias},
  title     = {Clustering and Summarization of Chat Dialogues},
  school    = {Link{\"o}ping University},
  year      = {2021},
  type      = {Master's thesis}
}

@inproceedings{hierarchicalsumm2014,
  author    = {Christensen, Janara and Soderland, Stephen and Bansal, Gagan and Mausam},
  title     = {Hierarchical Summarization: Scaling Up Multi-Document Summarization},
  booktitle = {Proceedings of the 52nd Annual Meeting of the Association for Computational Linguistics (ACL)},
  pages     = {902--912},
  year      = {2014},
  publisher = {Association for Computational Linguistics},
  doi       = {10.3115/v1/P14-1085}
}

@article{sharma2025kmeans,
  author    = {Sharma, Sanjay and others},
  title     = {Reducing Redundancy in Software Testing: A {K}-Means Clustering Approach to Test Case Minimization},
  journal   = {Journal of Information Systems Engineering and Management},
  volume    = {10},
  number    = {3s},
  year      = {2025},
  doi       = {10.55267/iadt.07.15669}
}

@inproceedings{clusterstructure2014,
  author    = {Duan, Chuan and Cleland-Huang, Jane},
  title     = {Clustering Support for Automated Tracing},
  booktitle = {Proceedings of the 22nd IEEE/ACM International Conference on Automated Software Engineering (ASE)},
  pages     = {244--253},
  year      = {2007},
  publisher = {ACM},
  doi       = {10.1145/1321631.1321668}
}

@article{conflictdetection2022,
  author    = {Mohamad, Sallam and Cailliau, Antoine and Darimont, Robert},
  title     = {Identifying the Requirement Conflicts in {SRS} Documents Using Sentence Transformers and {NER}},
  journal   = {arXiv preprint arXiv:2206.13690},
  year      = {2022}
}

@article{contractsumm2023,
  author    = {Jain, Chirag and Raje, Sushant and Deshpande, Girish},
  title     = {A Transformer-based Approach for Abstractive Summarization of Requirements from Obligations in Software Engineering Contracts},
  journal   = {Automated Software Engineering},
  volume    = {30},
  number    = {2},
  year      = {2023},
  publisher = {Springer},
  doi       = {10.1007/s10515-023-00399-5}
}

@article{reposummary2025,
  author    = {Zhang, Jingyi and Hou, Wanjun and Tang, Xin and Chen, Junhao and Zhou, Yuxin and Wei, Qing},
  title     = {{RepoSummary}: Feature-Oriented Summarization and Documentation Generation for Code Repositories},
  journal   = {arXiv preprint arXiv:2510.11039},
  year      = {2025}
}

@techreport{aspice2017,
  author    = {{VDA QMC Working Group 13}},
  title     = {Automotive {SPICE} Process Assessment / Reference Model, Version 3.1},
  institution = {Verband der Automobilindustrie (VDA)},
  year      = {2017}
}

@standard{iso26262,
  author    = {{International Organization for Standardization}},
  title     = {{ISO}~26262: Road Vehicles -- Functional Safety},
  year      = {2018},
  edition   = {2nd},
  note      = {Parts 1--12}
}

@article{clusterlabeling2019,
  author    = {Mund, Juergen and Femmer, Henning and Mendez, Daniel and Eckhardt, Jonas},
  title     = {Enhancing Software Requirements Cluster Labeling Using {Wikipedia}},
  journal   = {IEEE Access},
  volume    = {7},
  pages     = {145406--145419},
  year      = {2019},
  doi       = {10.1109/ACCESS.2019.2944610}
}

@article{graphclustering2023,
  author    = {Pudlitz, Florian and Brokhausen, Florian and Vogelsang, Andreas},
  title     = {Use Case Identification of Natural Language System Requirements with Graph-Based Clustering},
  journal   = {Design Science},
  volume    = {9},
  year      = {2023},
  publisher = {Cambridge University Press},
  doi       = {10.1017/dsj.2023.5}
}

@article{frontiers2025llmre,
  author    = {Siavashi, Fatemeh and Truscan, Dragos and Granmo, Ole-Christoffer},
  title     = {Research Directions for Using {LLM} in Software Requirement Engineering: A Systematic Review},
  journal   = {Frontiers in Computer Science},
  volume    = {7},
  year      = {2025},
  publisher = {Frontiers},
  doi       = {10.3389/fcomp.2025.1519437}
}

@inproceedings{ester1996dbscan,
  author    = {Ester, Martin and Kriegel, Hans-Peter and Sander, J{\"o}rg and Xu, Xiaowei},
  title     = {A Density-Based Algorithm for Discovering Clusters in Large Spatial Databases with Noise},
  booktitle = {Proceedings of the 2nd International Conference on Knowledge Discovery and Data Mining (KDD)},
  pages     = {226--231},
  year      = {1996},
  publisher = {AAAI Press}
}

@article{lewis2020bart,
  author    = {Lewis, Mike and Liu, Yinhan and Goyal, Naman and Ghazvininejad, Marjan and Mohamed, Abdelrahman and Levy, Omer and Stoyanov, Veselin and Zettlemoyer, Luke},
  title     = {{BART}: Denoising Sequence-to-Sequence Pre-training for Natural Language Generation, Translation, and Comprehension},
  booktitle = {Proceedings of the 58th Annual Meeting of the Association for Computational Linguistics (ACL)},
  pages     = {7871--7880},
  year      = {2020},
  doi       = {10.18653/v1/2020.acl-main.703}
}

@article{pure2017,
  author    = {Ferrari, Alessio and Dell'Orletta, Felice and Esuli, Andrea and Gervasi, Vincenzo and Gnesi, Stefania},
  title     = {Natural Language Requirements Processing: A 4D Vision},
  journal   = {IEEE Software},
  volume    = {34},
  number    = {6},
  pages     = {28--35},
  year      = {2017},
  doi       = {10.1109/MS.2017.4121207},
  note      = {PURE dataset available at \url{https://zenodo.org/records/1414117}}
}

@article{masuda2025ghl,
  author    = {Masuda, Satoshi and others},
  title     = {Generating High-Level Test Cases from Requirements using {LLM}: An Industry Study},
  journal   = {arXiv preprint arXiv:2510.03641},
  year      = {2025}
}

@inproceedings{adabala2025multistep,
  author    = {Adabala, Sai and others},
  title     = {Multi-Step Generation of Test Specifications using Large Language Models},
  booktitle = {Proceedings of the 63rd Annual Meeting of the Association for Computational Linguistics (ACL), Industry Track},
  year      = {2025},
  publisher = {Association for Computational Linguistics}
}

@inproceedings{chow2026automotive,
  author    = {Chow, Mo-Yuen and Kitani, Masanori},
  title     = {Testing Procedure Generation Based on Testing Requirements for Automotive Components with {LLM} Fine-Tuning},
  booktitle = {Natural Language Processing and Information Systems (NLDB 2025)},
  series    = {Lecture Notes in Computer Science},
  volume    = {15837},
  year      = {2026},
  publisher = {Springer},
  doi       = {10.1007/978-3-031-97144-0\_18}
}

@article{wang2025automotive,
  author    = {Wang, Song and Yu, Yevhen and Feldt, Robert and Parthasarathy, Dharani},
  title     = {Automating a Complete Software Test Process Using {LLMs}: An Automotive Case Study},
  journal   = {arXiv preprint arXiv:2502.04008},
  year      = {2025}
}

@article{koziolek2024control,
  author    = {Koziolek, Heiko and Ashiwal, Virendra and Bandyopadhyay, Somenath and Chandrika, K. R.},
  title     = {Automated Control Logic Test Case Generation using Large Language Models},
  journal   = {arXiv preprint arXiv:2405.01874},
  year      = {2024}
}

@article{clusteringtaxonomy2023,
  author    = {Ezugwu, Absalom E. and Ikotun, Abiodun M. and Oyelade, Olaide N. and Abualigah, Laith and Agushaka, Jeffery O. and Eke, Christopher I. and Akinyelu, Andronicus A.},
  title     = {A Comprehensive Survey of Clustering Algorithms: State-of-the-Art Machine Learning Applications, Taxonomy, Challenges, and Future Research Prospects},
  journal   = {Engineering Applications of Artificial Intelligence},
  volume    = {110},
  pages     = {104743},
  year      = {2022},
  doi       = {10.1016/j.engappai.2022.104743}
}
\end{document}